# A Microwave Oscillator Based on a Single Straintronic Magneto-tunneling Junction

Md Ahsanul Abeed, Justine L. Drobitch and Supriyo Bandyopadhyay[1]

Department of Electrical and Computer Engineering

Virginia Commonwealth University, Richmond, VA 23284, USA

There is growing interest in exploring nanomagnetic devices as potential replacements for electronic devices (e.g. transistors) in digital switching circuits and systems. A special class of nanomagnetic devices are switched with electrically generated mechanical strain leading to electrical control of magnetism. Straintronic magneto-tunneling junctions (s-MTJ) belong to this category. Their soft layers are composed of two-phase multiferroics comprising a magnetostrictive layer elastically coupled to a piezoelectric layer. Here, we show that a *single* straintronic magneto-tunneling junction with a passive resistor can act as a microwave oscillator whose traditional implementation would have required microwave operational amplifiers, capacitors and resistors. This reduces device footprint and cost, while improving device reliability. This is an analog application of magnetic devices where magnetic interactions (interaction between the shape anisotropy, strain anisotropy, dipolar coupling field and spin transfer torque in the soft layer of the s-MTJ) are exploited to implement an oscillator with reduced footprint.

## I.    INTRODUCTION

Straintronics is the field of switching the magnetization of a magnetostrictive nanomagnet, elastically coupled to an underlying piezoelectric film, with mechanical strain generated by applying a small voltage across the piezoelectric. This phenomenon can be exploited to realize extremely energy-efficient binary switches and a host of other digital systems that dissipate miniscule amounts of energy to operate [1]. What is less known, however, is that these same systems have another striking advantage, namely, their device characteristics may allow for combining the functionalities of multiple traditional devices into one single element. That would reduce device count significantly in both analog and digital circuits. Recently, it was shown theoretically that a *single* straintronic magneto-tunneling junction (s-MTJ), whose soft layer is a two-phase multiferroic composed of a magnetostrictive layer and a piezoelectric layer, can implement a ternary content-addressable memory cell [2] which would normally require 16 transistors to implement [3]. A single s-MTJ can also implement restricted Boltzmann machines which are unsupervised learning models of computation suitable to extract features from high-dimensional data [4].

In this paper, we show that a *single* s-MTJ along with a passive resistor can implement a microwave oscillator based on the interplay between strain anisotropy, shape anisotropy, dipolar magnetic field, and spin transfer torque generated by the passage of spin polarized current through the soft layer. Electronic implementation of a microwave oscillator would require microwave operational amplifiers (consisting of several transistors), capacitors and resistors [5]. In contrast, a *single* s-MTJ (Fig. 1) can realize the same functionality with just a single passive resistor.

[1] Corresponding author. Email: sbandy@vcu.edu

Magneto-tunneling junctions have been used extensively in the past for microwave spin-torque nano-oscillators (STNO) [6] which rely on magnetization precession induced by a spin polarized current. The principle of operation of the device analyzed here is different and relies on a feedback between strain, dipole coupling, shape anisotropy and spin polarized current delivering spin transfer torque in the soft layer of an MTJ. Unlike STNOs which typically have quality factors (ratio of resonant frequency to bandwidth) of 10 or less [6], the device presented here theoretically has a quality factor exceeding 500 according to the results of our simulations. This is due to the narrow bandwidth which is caused by the fact that the oscillations are spectrally pure (almost sinusoidal). The high quality factor is a significant advantage in antenna applications because it leads to higher radiation efficiency.

## II. OPERATING PRINCIPLE

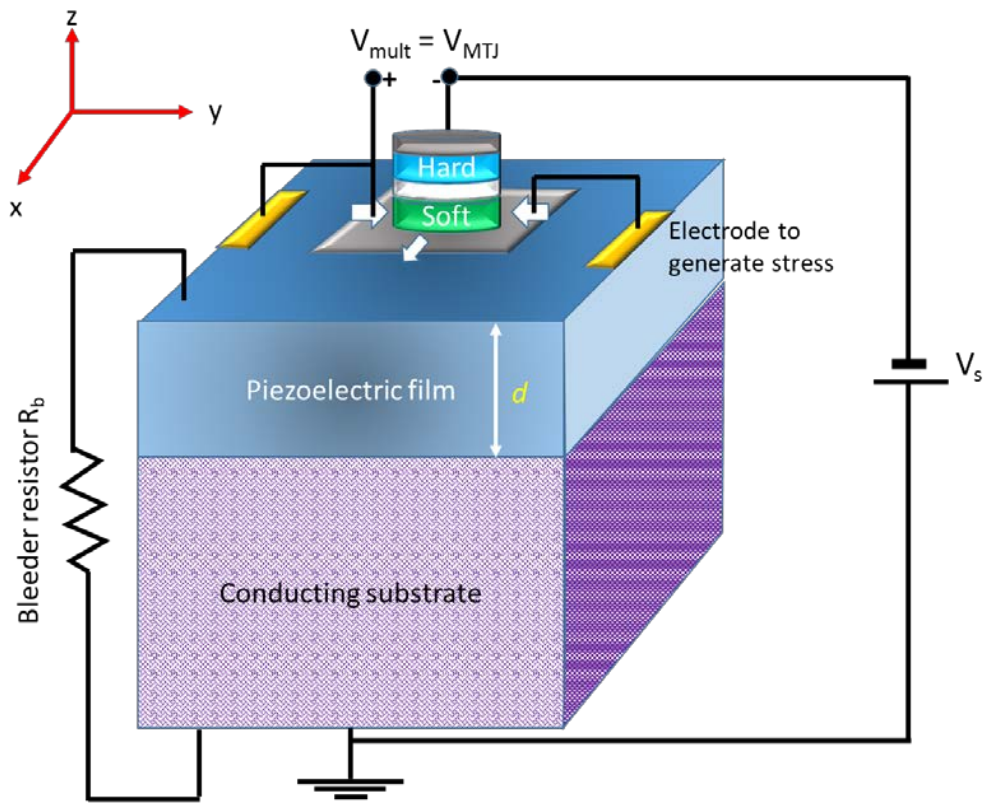

**Fig. 1**: A microwave oscillator implemented with a single straintronic magneto-tunneling junction (MTJ) and a passive resistor. The output voltage of the device is $V_{mult}$ which is the voltage dropped over the MTJ. The strain generated in the elliptical MTJ soft layer due to the voltage dropped over the piezoelectric is biaxial (compressive along the major axis and tensile along the minor axis). The white arrows show the strain directions. The piezoelectric layer is poled in the vertically up direction. The lateral dimension of the soft layer, the spacing between the edge of the soft layer and the nearest electrode, and the piezoelectric film thickness are all approximately the same and that generates biaxial strain in the soft layer [7].

To understand how a single s-MTJ can implement microwave oscillator, consider the s-MTJ in Fig. 1 whose hard and soft layers have in-plane magnetic anisotropy. The two layers are shaped into elliptical disks and their major axes (easy axes) are collinear. The hard layer is permanently magnetized in one direction along its major axis (along the +y-axis). It is made of a synthetic anti-ferromagnet and engineered to produce a weak (but non-zero) dipole coupling field on the soft layer. In this configuration, the dipole coupling field in the soft layer is constant and oriented along the latter's major axis in the direction opposite to the magnetization of the hard layer [8], i.e. along the –y-direction. Therefore, we can write the dipole coupling field as

$$\vec{H}_{\text{dipole}} = -H_d \hat{y} \qquad (1)$$

where $H_d$ is a constant that depends on the composition of the hard layer.

When no current is passing through the s-MTJ, the magnetizations of the hard and soft layers will be mutually antiparallel owing to the dipole coupling field and the MTJ will be in the high resistance state. If we now turn on the voltage supply $V_s$ (with the polarity shown in Fig. 1b), spin-polarized electrons will be injected from the hard layer into the soft layer and they will gradually turn the soft layer's magnetization in the direction of the hard layer's magnetization because of the generated spin transfer torque. This will take the MTJ toward the low resistance state.

Note that the voltage dropped over the piezoelectric film is

$$V_{piezo} = \frac{R_{piezo} \| R_b}{R_{piezo} \| R_b + R_{MTJ}} \qquad (2)$$

where $R_{piezo}$ is the resistance of the piezoelectric film between the s-MTJ soft layer and the conducting substrate, $R_b$ is the resistance of a current bleeder resistor in parallel with the piezoelectric film as shown in Fig. 1, and $R_{MTJ}$ is the resistance of the s-MTJ. The resistance of the conducting substrate is assumed to be negligible.

Equation (2) shows that when the s-MTJ goes into the low resistance state $(R_{MTJ} \to \text{low})$, the voltage dropped over the piezoelectric film $V_{piezo}$ increases and that generates sufficient strain in that film which is partially transferred to the soft layer. Since the soft layer is magnetostrictive, this strain will rotate its magnetization away from the major axis toward the minor axis because of the Villari effect, as long as the product of the strain and magnetostriction coefficient is negative (i.e. strain will have to be compressive if the magnetostriction coefficient of the soft layer is positive and tensile if the magnetostriction coefficient is negative). This rotation, which increases the angle between the magnetizations of the hard and soft layer, will increase the resistance of the MTJ and reduce the spin polarized current flowing through it (for a constant supply voltage $V_s$). At that point, the dipole coupling effect can overcome the reduced spin transfer torque associated with the reduced current and swing the soft layer's magnetization toward an orientation antiparallel to that of the hard layer, causing the MTJ resistance to approach the high resistance state. Once that happens, the voltage dropped over the piezoelectric film falls (see Equation (2) again) and the strain in the soft layer subsides. However, the spin polarized current still flows through the soft magnet, and over sufficient time, will transfer enough torque to the soft layer's magnetization to make it once again attempt to align parallel to the hard layer's magnetization. This will take the MTJ back to the low resistance state and the process repeats itself. The MTJ resistance $R_{MTJ}$ therefore continuously oscillates between the high and low states. This will make the voltage $V_{MTJ}$ dropped over the MTJ $\left[V_{MTJ} = V_s R_{MTJ} / \left( R_{piezo} \| R_b + R_{MTJ} \right)\right]$ also continuously oscillate between two values, resulting in an oscillator. This is operating principle of the s-MTJ based oscillator.

### III. EQUATIONS OF MAGNETO-DYNAMICS

In order to simulate the temporal dynamics of the oscillator, we have to simulate the magnetodynamics of the soft layer under the combined effects of strain anisotropy, shape anisotropy, dipolar coupling field and spin transfer torque. This will yield the time variations of $R_{MTJ}$ and $V_{MTJ}$, which is the output of the oscillator. To do this, we have solved the stochastic Landau-Lifshitz-Gilbert (LLG) equation, assuming that the magneto-dynamics of the soft layer can be described within the macrospin approximation. We understand that the soft layer can be multi-domain, but we avoided the more accurate stochastic micromagnetic simulations in the presence of thermal noise since it will overwhelm our computational resources and would produce only small quantitative differences, while the qualitative results will not change.

The LLG equation describing the magnetodynamics of the soft layer is:

$$\frac{d\vec{m}(t)}{dt} = -\gamma \vec{m}(t) \times \vec{H}_{total}(t) + \alpha \left( \vec{m}(t) \times \frac{d\vec{m}(t)}{dt} \right) + a\vec{m}(t) \times \left( \frac{\eta \vec{I}_s(t) \mu_B}{qM_s\Omega} \times \vec{m}(t) \right) + b \frac{\eta \vec{I}_s(t) \mu_B}{qM_s\Omega} \times \vec{m}(t) \quad (3)$$

where

$\hat{m}(t) = m_x(t)\hat{x} + m_y(t)\hat{y} + m_z(t)\hat{z} \qquad \left[ m_x^2(t) + m_y^2(t) + m_z^2(t) = 1 \right]$

$\vec{H}_{total} = \vec{H}_{demag} + \vec{H}_{stress} + \vec{H}_{dipole} + \vec{H}_{thermal}$

$\vec{H}_{demag} = -M_s N_{d-xx} m_x(t)\hat{x} - M_s N_{d-yy} m_y(t)\hat{y} - M_s N_{d-zz} m_z(t)\hat{z}$

$\vec{H}_{stress} = \dfrac{3}{\mu_0 M_s} \left( \lambda_s \sigma_{yy}(t) m_y(t) \right) \hat{y}$

$\vec{H}_{dipole} = -H_d \hat{y} \qquad [\text{constant independent of time}]$

$\vec{H}_{thermal} = \sqrt{\dfrac{2\alpha kT}{\gamma (1+\alpha^2) \mu_0 M_s \Omega (\Delta t)}} \left[ G^x_{(0,1)}(t)\hat{x} + G^y_{(0,1)}(t)\hat{y} + G^z_{(0,1)}(t)\hat{z} \right]$

The last term in the right hand side of Equation (3) is the field-like spin transfer torque and the second to last term is the Slonczewski torque. The coefficients $a$ and $b$ depend on device configurations and following [9], we will use the values $a=1$, $b=0.3$. Here $\hat{m}(t)$ is the time-varying magnetization vector in the soft layer normalized to unity, $m_x(t)$, $m_y(t)$ and $m_z(t)$ are its time-varying components along the x-, y- and z-axis, $\vec{H}_{demag}$ is the demagnetizing field in the soft layer due to shape anisotropy, $\vec{H}_{stress}$ is the magnetic field caused by stress in the magnetostrictive soft layer, $\vec{H}_{dipole}$ is the dipole coupling field given in Equation (1) and $\vec{H}_{thermal}$ is the random magnetic field due to thermal noise. The different parameters in Equation (3) are: $\gamma = 2\mu_B\mu_0/\hbar$ (gyromagnetic ratio), $\alpha$ is the Gilbert damping constant, $\mu_0$ is the magnetic permeability of free space, $M_s$ is the saturation magnetization of the magnetostrictive soft layer, $kT$ is the thermal energy, $\lambda_s$ is the magnetostriction coefficient of the soft layer, $\Omega$ is the volume of the soft layer which is given by $\Omega = (\pi/4)a_1a_2a_3$ [$a_1$ = major axis, $a_2$ = minor axis and $a_3$ = thickness], $\sigma_{yy}(t)$ is the uniaxial time-varying stress created in the soft layer by the voltage dropped over the piezoelectric film, $\Delta t$ is the time step used in the simulation, and $G^x_{(0,1)}(t)$, $G^x_{(0,1)}(t)$ and $G^x_{(0,1)}(t)$ are three uncorrelated Gaussians with zero mean and unit standard deviation [10]. The quantities $N_{d-xx}, N_{d-yy}, N_{d-zz}$ [$N_{d-xx} + N_{d-yy} + N_{d-zz} = 1$] are calculated from the dimensions of the elliptical soft layer following the prescription of ref. [11]. We assume that the charge current injected from the hard layer into the soft layer is $\vec{I}_s(t)$ and that the spin polarization in the current is $\eta$. The spin current is given by

$$\eta \vec{I}_s(t) = \eta |\vec{I}_s(t)| \hat{y} \text{ where } |\vec{I}_s(t)| = \frac{V_s}{R_{MTJ}(t) + R_{piezo} || R_b} = \frac{V_s}{R_P + \frac{\Delta R}{2}(1-m_y(t)) + R_{piezo} || R_b}.$$

(4)

Here, $R_P$ is the low resistance of the MTJ (when the magnetizations of the hard and soft layers are closest to being parallel) and $R_{AP}$ is the high resistance when the two magnetizations are closest to being anti-parallel. The quantity $\Delta R = R_{AP} - R_P$. The stress generated in the soft layer is actually biaxial (compressive along the major axis and tensile along the minor axis for the power supply voltage polarity shown in Fig. 1, as long as the piezoelectric is poled in the vertically down direction) [7]. The conditions for such stress generation, discussed extensively in ref. [7], are stated in the caption of Fig. 1. Since the effect of biaxial stress is difficult to incorporate in the LLG equation, we have approximated the stress as uniaxial along the major axis. Note that tensile stress along the minor axis has the same effect on magnetization as compressive stress along the major axis (and vice versa); hence, we can combine the two effects into the effect of a single uniaxial stress along the major axis. The uniaxial strain generated in the piezoelectric is the product of the electric field in the piezoelectric film along its thickness (generated by $V_{piezo}$) and its $d_{31}$ coefficient, and we assume that it is fully transferred to the soft layer. The uniaxial stress in the soft layer is therefore given by

$$\sigma_{yy}(t) = d_{31}Y\frac{R_{piezo} || R_b}{R_{piezo} || R_b + R_{MTJ}(t)}\frac{V_s}{d} = d_{31}Y\frac{R_{piezo} || R_b}{R_{piezo} || R_b + R_P + \frac{\Delta R}{2}(1-m_y(t))}\frac{V_s}{d}, \quad (5)$$

where $d$ is the thickness of the piezoelectric film and $Y$ is the Young's modulus of the soft layer material. Finally, the voltage drop over the s-MTJ, is the *oscillator output voltage* and it is given by

$$V_{oscillator}(t) = V_{MTJ}(t) = V_s\frac{R_{MTJ}(t)}{R_{piezo} || R_b + R_{MTJ}(t)} = V_s\frac{R_P + (\Delta R/2)(1-m_y(t))}{R_{piezo} || R_b + R_P + (\Delta R/2)(1-m_y(t))}. \quad (6)$$

We will assume a tunneling magnetoresistance ratio (TMR) of 500% [12], which will make $\Delta R = 5R_P$.

Using the vector identity $\vec{a} \times (\vec{b} \times \vec{c}) = \vec{b}(\vec{a} \cdot \vec{c}) - \vec{c}(\vec{a} \cdot \vec{b})$, we can recast the vector equation in Equation (3) as

$$(1+\alpha^2)\frac{d\vec{m}(t)}{dt} = -\gamma\left(\vec{m}(t) \times \vec{H}_{total}(t)\right) - \gamma\alpha\left[\vec{m}(t)\left(\vec{m}(t) \cdot \vec{H}_{total}(t)\right) - \vec{H}_{total}(t)\right]$$
$$-(\alpha a - b)\left(\frac{\eta \vec{I}_s(t)\mu_B}{qM_s\Omega} \times \vec{m}(t)\right) + (a+\alpha b)\frac{\eta I_s(t)\mu_B}{qM_s\Omega}\left[\hat{y} - \vec{m}(t)m_y(t)\right] \quad (7)$$

This vector equation can be recast as three coupled scalar equations in the three Cartesian components of the magnetization vector:

$$(1+\alpha^2)\frac{dm_x(t)}{dt} = -\gamma\left[m_y(t)H_z(t) - m_z(t)H_y(t)\right] - \alpha\gamma\left[m_x(t)\left[m_x(t)H_x(t) + m_y(t)H_y(t) + m_z(t)H_z(t)\right] - H_x(t)\right]$$
$$-(a\alpha - b)\frac{\eta I_s(t)m_z(t)\mu_B}{qM_s\Omega} - (a+\alpha b)\frac{\eta I_s(t)m_x(t)m_y(t)\mu_B}{qM_s\Omega}$$

$$(1+\alpha^2)\frac{dm_y(t)}{dt} = -\gamma\left[m_z(t)H_x(t) - m_x(t)H_z(t)\right] - \alpha\gamma\left[m_y(t)\left[m_x(t)H_x(t) + m_y(t)H_y(t) + m_z(t)H_z(t)\right] - H_y(t)\right], \quad (8)$$
$$+(a+\alpha b)\frac{\eta I_s(t)\mu_B}{qM_s\Omega}(1 - m_y^2(t))$$

$$(1+\alpha^2)\frac{dm_z(t)}{dt} = -\gamma\left[m_x(t)H_y(t) - m_y(t)H_x(t)\right] - \alpha\gamma\left[m_z(t)\left[m_x(t)H_x(t) + m_y(t)H_y(t) + m_z(t)H_z(t)\right] - H_z(t)\right]$$
$$+(a\alpha - b)\frac{\eta I_s(t)m_x(t)\mu_B}{qM_s\Omega} - (a+\alpha b)\frac{\eta I_s(t)m_z(t)m_y(t)\mu_B}{qM_s\Omega}$$

where

$$H_x = -M_s N_{d-xx} m_x(t) + \sqrt{\frac{2\alpha kT}{\gamma(1+\alpha^2)\mu_0 M_s \Omega(\Delta t)}} G^x_{(0,1)}(t)$$
$$H_y = -M_s N_{d-yy} m_y(t) + \sqrt{\frac{2\alpha kT}{\gamma(1+\alpha^2)\mu_0 M_s \Omega(\Delta t)}} G^y_{(0,1)}(t) + \frac{3}{\mu_0 M_s}\left(\lambda_s \sigma_{yy}(t) m_y(t)\right)\hat{y} - H_d\hat{y} \quad (9)$$
$$H_z = -M_s N_{d-zz} m_z(t) + \sqrt{\frac{2\alpha kT}{\gamma(1+\alpha^2)\mu_0 M_s \Omega(\Delta t)}} G^z_{(0,1)}(t)$$

The quantity $\Delta t$ is the time step used in solving Equation (8) numerically. We solve Equation (8) numerically to find the magnetization component $m_y(t)$ as a function of time and then use Equation (6) to find $V_{oscillator}(t)$ as a function of time. In our simulation, we take $\Delta t = 0.1$ ps and we have verified that a smaller value does not change anything perceptibly.

We assumed that the soft layer of the MTJ is made of Terfenol-D and used the material and device parameters in Table I. The hard layer (fashioned out of a synthetic antiferromagnet, such as FeCoB with

Rh spacer layers [13]) has the same geometry as the soft layer. The adhesion layer at the bottom of the MTJ stack is assumed to be thin enough to not impede strain transfer from the underlying piezoelectric into the soft layer. The top layer above the hard layer is intended to prevent oxidation of the hard layer.

**Table I: Material and device parameters**

| Saturation magnetization ($M_s$) | $8 \times 10^5$ A/m |
|---|---|
| Magnetostriction coefficient ($\lambda_s$) | $600 \times 10^{-6}$ |
| s-MTJ resistance in the parallel state ($R_p$) | 13.33 ohms |
| Piezoelectric film resistance in parallel with the bleeder resistance $\left( R_{piezo} \parallel R_b \right)$ | 25.4 ohms |
| Elliptical soft layer's major axis ($a_1$) | 800 nm |
| Elliptical soft layer's minor axis ($a_2$) | 700 nm |
| Elliptical soft layer's thickness ($a_3$) | 2.2 nm |
| Supply voltage ($V_s$) | 12 V |
| Gilbert damping constant ($\alpha$) | 0.01 |
| Spin polarization in the current ($\eta$) | 0.3 |
| Piezoelectric film thickness ($d$) | 1 µm |
| Young's modulus of soft layer ($Y$) | 80 GPa |
| Piezoelectric coefficient ($d_{31}$) | 300 pm/V |
| Dipole coupling field ($H_d$) | 7957 A/m (10 mT) |

## IV. RESULTS

We carried out stochastic simulation of the magnetodynamics of the soft layer (Equation (8)) assuming that the supply voltage $V_s$ is turned on at time $t = 0$. The initial orientation of the magnetization at $t = 0$ is close to the –y-direction because of dipole coupling with the hard layer whose magnetization is in the +y-direction. We therefore take the initial values of the soft layer's magnetization components as $m_x(t=0) = 0.045, m_y(t=0) = -0.999, m_z(t=0) = 0.001$. We do not start with $m_y(t=0) = -1$ since that is a stagnation point where the torque on the magnetization due to the spin polarized current, shape anisotropy, strain and dipole coupling, vanishes. The thermal noise field can dislodge the magnetization from the stagnation point. At 0 K, there is no thermal noise and hence the magnetization would have been stuck at –y-axis had we used the initial condition $m_y(t=0) = -1$.

In Fig. 2, we show the temporal variations of the magnetization components $m_x(t), m_y(t), m_z(t)$ at 0 K temperature when there is no thermal noise and the dipole coupling field strength $H_d$ is 7957 A/m (10 mT). This figure shows that the magnetization oscillates by nearly the full $180^0$ (between the parallel and anti-parallel configurations) since $m_y(t)$ varies between -0.9394 and 0.9009. Note also that as expected, the magnetization lifts out of the soft layer's plane during the oscillation since $m_z(t)$ varies between -0.52 and 0.52.

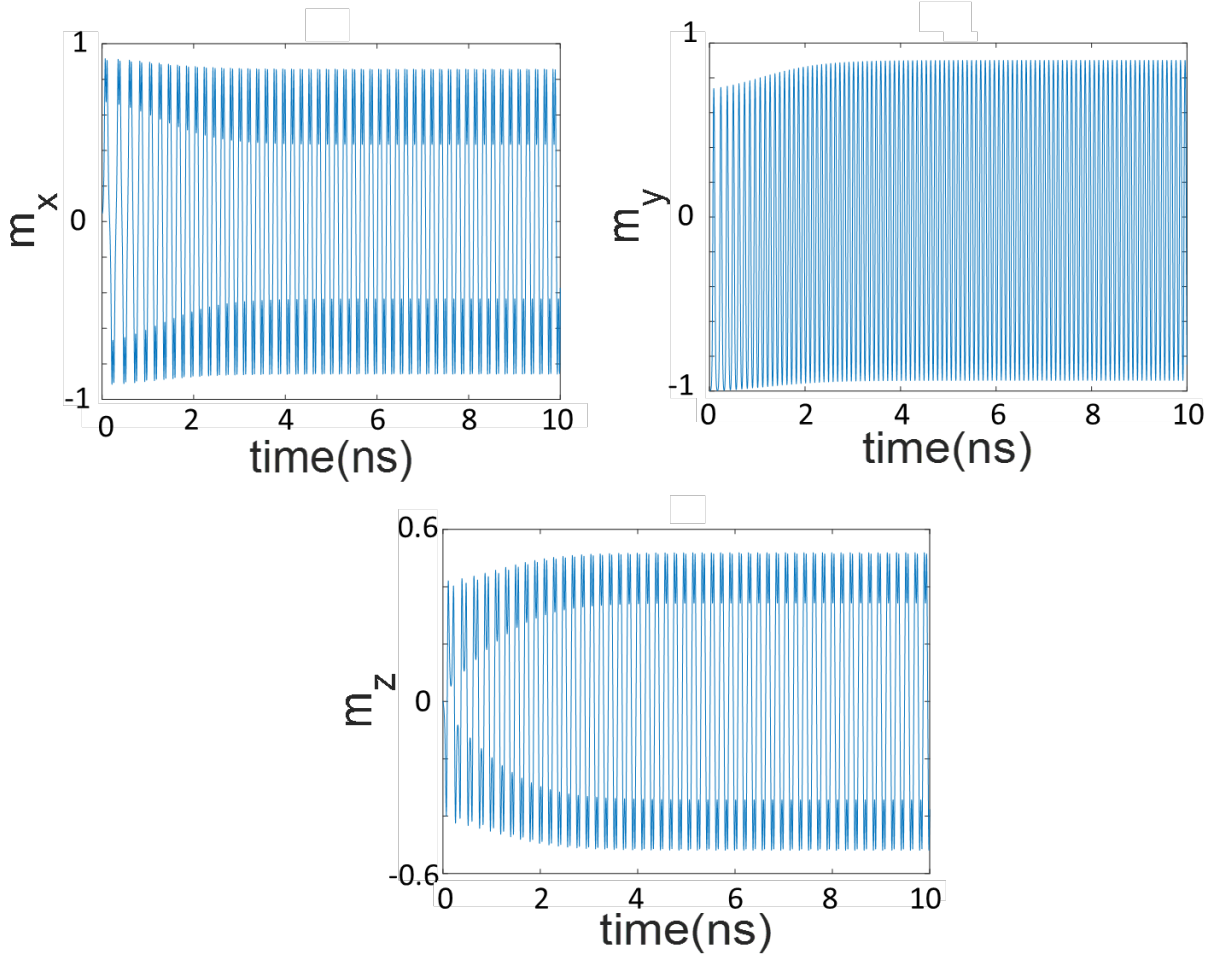

Fig. 2: Temporal variations of the magnetization components in the soft layer at 0 K temperature. Note that it takes almost 3 ns to reach steady state amplitude.

In Fig. 3, we show how $R_{MTJ}$ varies with time at 0 K. Note that the maximum and minimum values of $R_{MTJ}$ do not vary by a factor of 6 (despite the assumed TMR of 500%) since the magnetization of the soft layer never reaches either the full antiparallel direction or the full parallel direction during its swing ($m_y(t)$ varies between -0.9394 and 0.9009). The maximum and minimum values of the s-MTJ resistance differ by a factor of ~4.7, instead of 6.

In Fig. 4, we show how the charge current flowing through the MTJ varies with time. The peak current density is $6.5 \times 10^{11}$ A/m$^2$ which is high, but not completely unreasonable since current densities of this order have been used in MTJs [6, 14, 15]. Reducing it would have required reducing the power supply voltage $V_s$. Unfortunately, a lower power supply voltage is not able to sustain the oscillations. Moreover, in the presence of noise at room temperature, random fluctuations in the thermally averaged oscillation amplitude show up if the power supply voltage is lower than a minimum threshold value. The threshold voltage is determined by the complex interaction between the shape anisotropy, strain anisotropy, STT and dipole coupling field. For the parameters that we chose for the s-MTJ, the threshold voltage is 12 V, which results in the current density is $6.5 \times 10^{11}$ A/m$^2$. Any lower current density will not sustain the oscillations or be able to counter the effect of room temperature thermal noise.

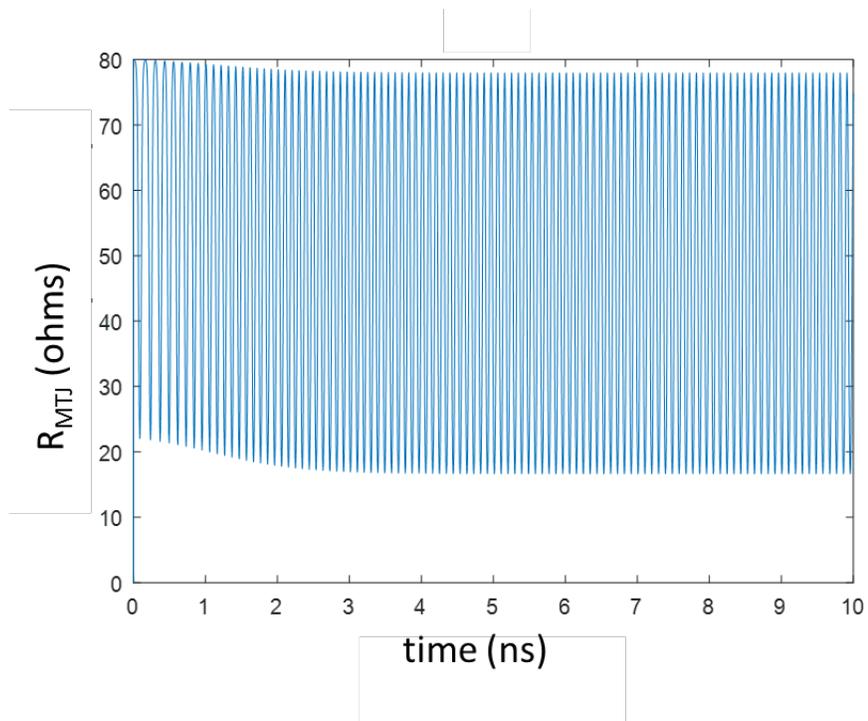

Fig. 3: Variation of the resistance of the s-MTJ with time at 0 K temperature. Again, it takes almost 3 ns to reach steady state.

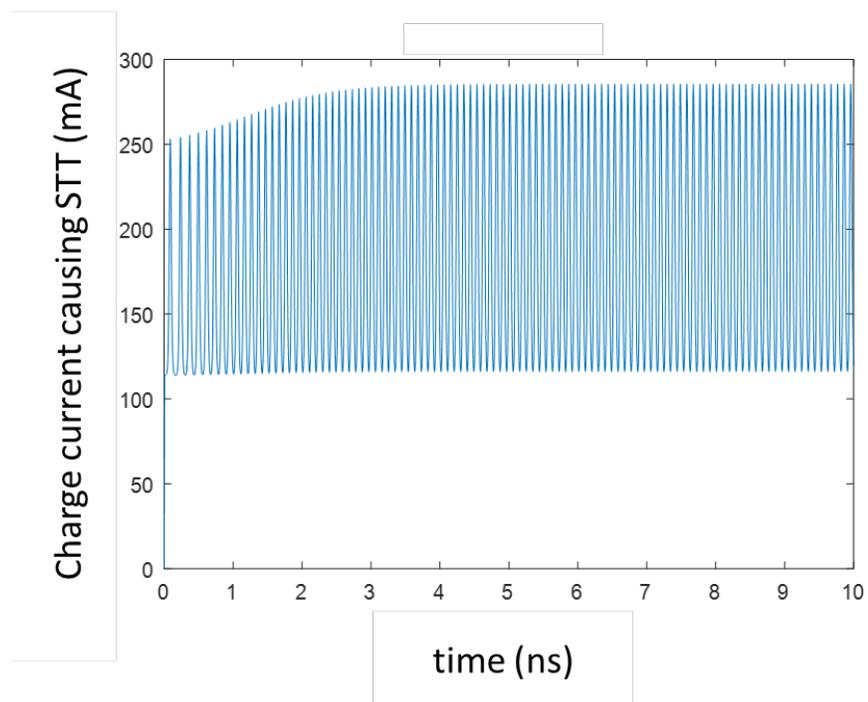

Fig. 4: Variation of the current through the s-MTJ with time at 0 K. This current delivers the spin transfer torque. Again, it takes almost 3 ns to reach steady-state amplitude.

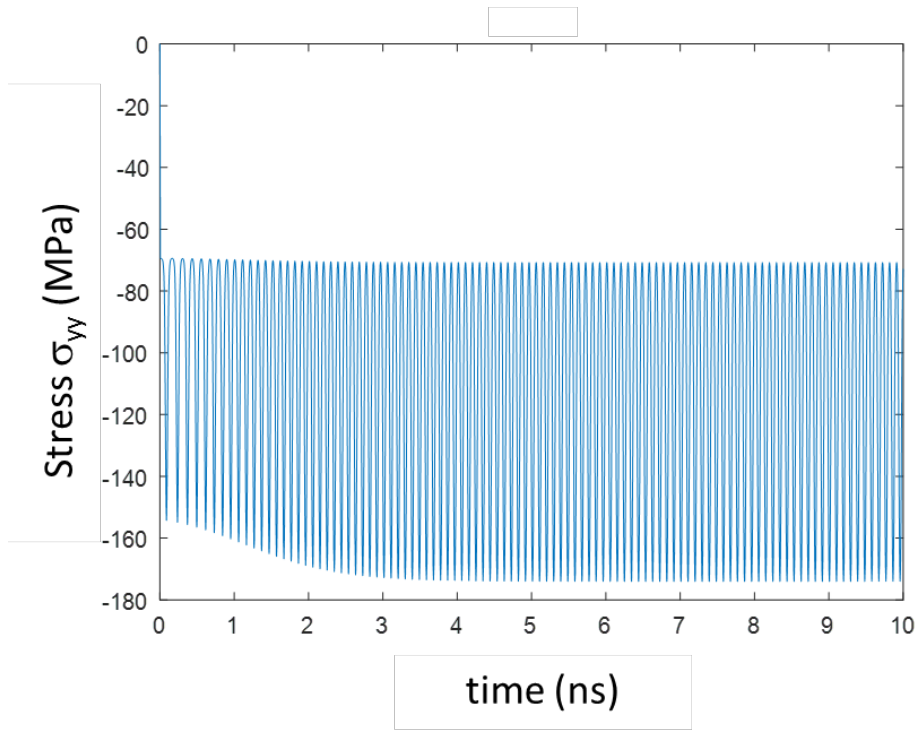

Fig. 5: Variation of the uniaxial stress in the soft layer as a function of time at 0 K. It takes about 3 ns to reach steady-state amplitude.

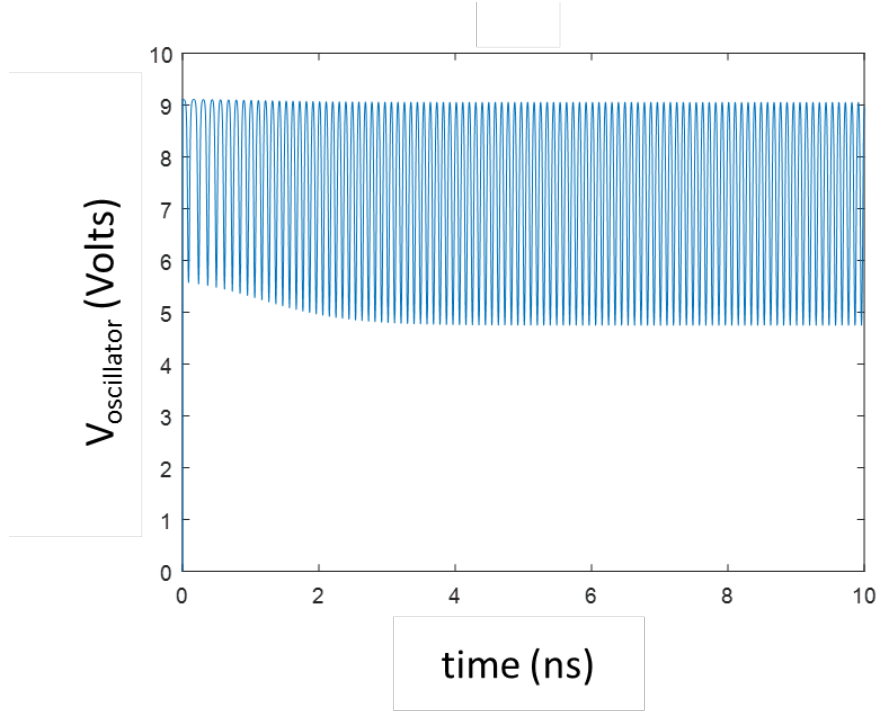

Fig. 6: Variation of the voltage across the s-MTJ (the oscillator output voltage) as a function of time at 0 K. It takes about 3 ns to reach steady-state amplitude. There is a dc offset of ~7.4 V, while the oscillatory peak-to-peak amplitude is slightly over 4.3 V.

One likely approach to reduce the threshold current density is to increase the area of the s-MTJ while keeping the soft layer's volume constant. This does decrease the threshold current density somewhat, but not in inverse proportion to the area. We chose the s-MTJ surface area $[(\pi/4)\times 700 \text{ nm} \times 800 \text{ nm}]$ such that it remains sub-micron in lateral dimensions, but not so small as to make the threshold current density too high. Micron scale MTJs have been fabricated by many groups [16, 17].

In Fig. 5, we show how the uniaxial stress $\sigma_{yy}(t)$ along the major axis of the soft layer varies with time – all at 0 K when there is no thermal noise. Note that the maximum compressive stress generated in the soft layer is about 175 MPa. Since the Young's modulus of Terfenol D is 80 GPa, the maximum strain generated in the soft layer is 0.2%. There are reports of 0.6% strain generation in piezoelectric materials [18]; hence, this value is reasonable.

Finally, in Fig. 6, we show how the oscillator output voltage $V_{\text{oscillator}}(t)$ varies with time at 0 K.

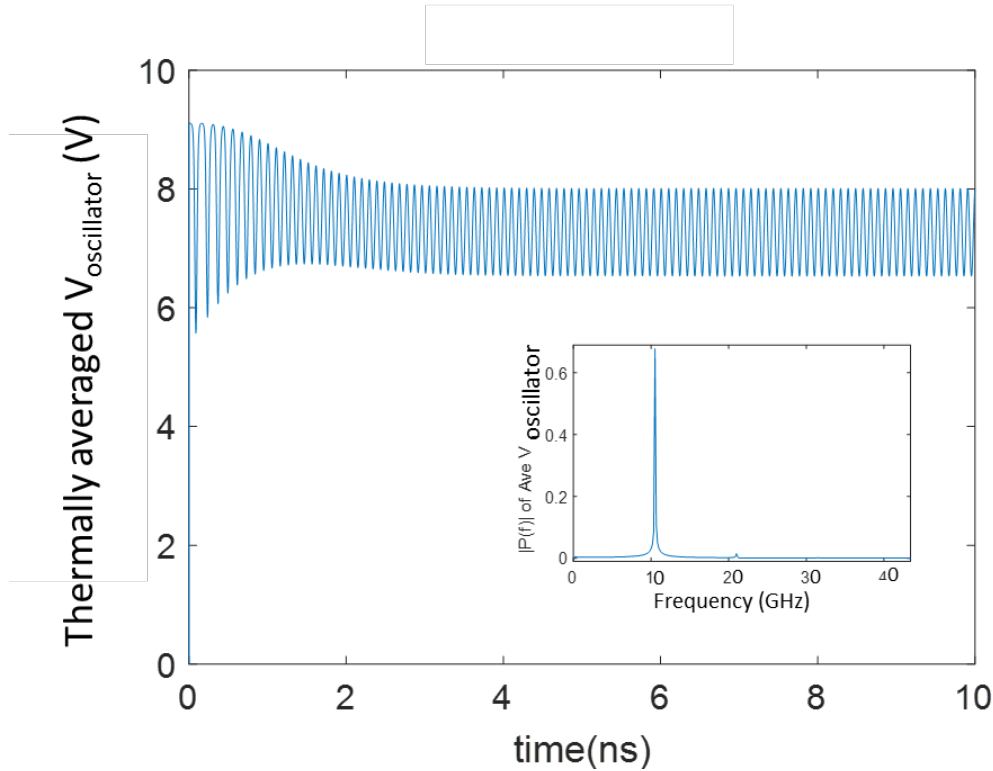

Fig. 7: Thermally averaged variation in the oscillator voltage as a function of time at 300 K in the presence of thermal noise. This plot has been obtained by averaging 1000 trajectories (results of simulations) which are all slightly different from each other because of the randomness of the noise field. It again takes about 3 ns to reach steady state amplitude. The steady state period is ~100 ps (frequency = 10.52 GHz, wavelength = 3 mm). The dc offset is still about 7.3 V and the steady state peak-to-peak amplitude is 1.5 V. The inset shows the Fourier spectra of the oscillations after suppressing the dc component. The fundamental frequency is 10.52 GHz and there is a second harmonic at ~21 GHz whose amplitude is ~60 times less than that of the fundamental. Surprisingly the output is spectrally pure and this is almost a monochromatic (ideal) oscillator. The resonant frequency is 10.52 GHz and the bandwidth (full width at half maximum) is ~200 MHz, leading to a quality factor of 526.

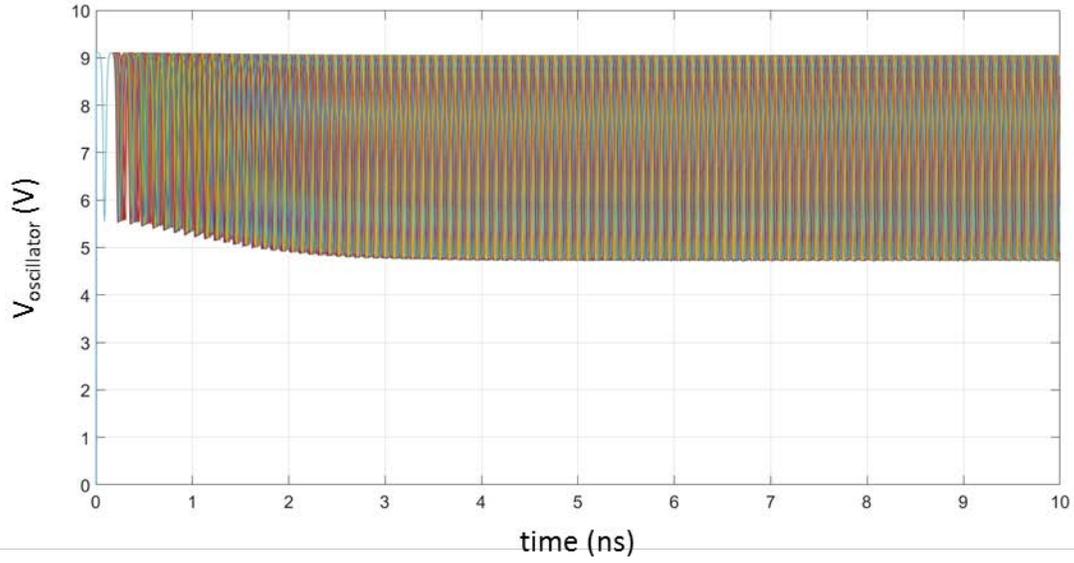

Fig. 8: Oscillator voltage as a function of time at 300 K. Here, 1000 different trajectories have been plotted and clearly there is some spread among the trajectories because of thermal noise. The plot in Fig. 7 was obtained by averaging over these trajectories.

In Fig. 7, we show the "average" $V_{oscillator}(t)$ as a function of time at 300 K in the presence of thermal noise. The averaging is carried out over 1000 trajectories (simulation results) shown in Fig. 8, each trajectory is slightly different from the other owing to random thermal noise. Note from Fig. 7 that thermal noise does not inhibit the oscillator operation, but introduces a very slight fluctuation in the amplitude. To understand the origin of this fluctuation, we have plotted the oscillations of the 1000 different trajectories in Fig. 8. Note that the different trajectories are slightly phase-shifted from each other, which is what leads to the smearing. Hence, averaging over them causes a slight random fluctuation in the amplitude of the oscillation seen in Fig. 7. The fluctuation is however less than 5% of the amplitude and hence not inhibitory.

Note that in all cases, it takes about 3 ns to reach steady-state amplitude. This is the time taken by the complex interaction between shape anisotropy, stress anisotropy, dipole field and spin transfer torque to reach steady state.

We found that the oscillator operation is sensitive to the choice of various parameters such as the supply voltage $V_s$, the dipole coupling field $H_d$, $R_{piezo} \parallel R_b$ and $R_p$. There are certain "windows" for these quantities where the oscillations show up. Outside these windows, the oscillations are suppressed. This is not surprising. The phenomenon underlying the operation of this device depends on a delicate balance between the magnetodynamics due to STT current, dipole coupling field, shape anisotropy (demagnetizing) field and strain, and the complex *interaction* between them. The STT current tries to align the magnetization along the soft layer's major axis in the direction of the hard layer's magnetization, the strain tries to align the magnetization perpendicular to the major axis, while the dipole coupling field tries to align the magnetization along the major axis anti-parallel to the direction of the hard layer's magnetization. Thus, the three different agents attempt to align the magnetization along three different directions and it is the complex interaction between these agents that cause the magnetization direction to oscillate in time. Clearly, too high a stress will pin the magnetization permanently along the minor axis, too high STT will pin the magnetization parallel to the hard layer's magnetization and too high dipole coupling will pin the magnetization anti-parallel to the hard layer's magnetization. Only the right balance between these agents will make the magnetization oscillate. Because of the need for this delicate balance, the oscillator operation is sensitive to the choice of device parameters.

The sensitivity is not a serious impediment since the device parameters like the dipole coupling field and the power supply voltage can be easily tuned and hence calibrated. The dipole field acts as a constant field along the major axis of the soft layer and is directed opposite to the magnetization of the hard layer. Therefore, this field can be easily tuned with an external in-plane magnetic field directed along the major axis of the soft layer. Similarly, the power supply voltage can also be tuned and calibrated.

The dipole coupling field strength is given by

$$H_d = \xi \frac{M_s \Omega}{4\pi r^3}, \qquad (10)$$

where $r$ is the center-to-center separation between the hard and soft layers along the z-axis and $\xi$ is a suppression factor $(0 \leq \xi \leq 1)$ due to the use of a synthetic antiferromagnet for the hard layer, which suppresses dipole interaction. We assume that the spacer layer of the MTJ is 2 nm thick and the hard layer is 10 nm thick (because it consists of multiple layers), which would make $r = 8.1$ nm. Using the values in Table I for $M_s$ and $\Omega$, we find that for $H_d$ to be 7957 A/m (10 mT), $\xi \approx 4.3 \times 10^{-6}$. Again, there is no criticality associated with the value of $\xi$ since the dipole coupling field can be tuned by an in-plane magnetic field directed along the major axis of the elliptical soft layer.

The power dissipated in the oscillator is mostly static (dc) power and is ~ 182 mA $\times$ 12 V = 2.18 W. The rms ac power output is ~ 30 mW. A single oscillator device that outputs 30 mW at microwave frequency is quite attractive.

## V. CONCLUSION

In conclusion, we have shown that a single straintronic magneto-tunneling junction (with a passive resistor) can implement a microwave oscillator with excellent quality factor which would have taken multiple electronic components to implement in traditional electronics. In particular, one would have needed microwave components which are expensive. Thus, the magnetic implementation has several advantages, such as lower cost and reduced device footprint. The operation is sensitive to the choice of the supply voltage and the dipolar coupling field for any given MTJ specification, but these quantities can be easily tuned with external agents to calibrate a prototype for optimal operation.

**Acknowledgement:** This work is partially supported by the US National Science Foundation under grants ECCS-1609303 and CCF-1815033.